\documentstyle[aps,epsf]{revtex}
\begin{document}

\twocolumn [
\hsize\textwidth\columnwidth\hsize\csname@twocolumnfalse\endcsname

\title{Thermal dephasing and 
the echo effect in a confined  Bose-Einstein condensate }
\author{A. B. Kuklov, and  N. Chencinski}
\address{ Department of Applied Sciences,
The College of  Staten Island, CUNY,
     Staten Island, NY 10314}

\maketitle
\begin{abstract}

It is shown that thermal fluctuations of the normal component
induce dephasing -- reversible damping of 
 the low energy
collective modes of a confined Bose-Einstein  
condensate. The dephasing rate
is calculated for the isotropic oscillator trap, where
Landau damping is expected to be suppressed.
This rate is characterized by a steep temperature 
dependence, and it is weakly 
amplitude dependent. In the limit of large numbers
of bosons forming the condensate, the rate approaches zero.
 However, for the numbers employed by 
the JILA group, the calculated value of the rate is close  
to the experimental one. We suggest 
that a reversible nature of the damping caused by the thermal
dephasing in the isotropic trap can be tested by the echo effect.
A reversible nature of Landau damping  
is also discussed,
and a possibility of observing the echo effect
 in an anisotropic trap  
is considered as well. 
The parameters of the echo 
are calculated in the weak echo limit for the isotropic trap.
 Results of the numerical
simulations of the echo are also presented. 
\\

\end{abstract}
\vskip0.5 cm
]

\section{Introduction}
Properties of the trapped ultracold atomic gases demonstrating 
the phenomenon of Bose-Einstein condensation 
\cite{COR,KET,HUL} attract much of attention.
A confined geometry of the cloud leads to circumstances
under which new phenomena such as, e.g., a specific scaling in the
thermodynamical properties \cite{PIT1} and the two step condensation
\cite{TWO} can be anticipated. It was also suggested that the phase of 
the confined condensate should demonstrate the phenomenon of
collapses and revivals \cite{COLL} and the phase diffusion effect
\cite{YOU}.

Recently  it has been discussed in Refs. \cite{KUK,PIT2},
 that the dynamical quantum  behavior of the
atomic cloud consisting of a finite number of bosons is very different
from the properties of infinite systems. Specifically, an amplitude of
a normal mode of the confined condensate should demonstrate quantum
dephasing which results in an apparent damping of the mode even at zero
temperature $T$. The rate of such damping is determined by the 
interparticle interaction and turns out to be linearly dependent on
the mode amplitude, if it is small \cite{KUK,PIT2}. A possibility
of the mode revival has also been predicted \cite{KUK,PIT2}. 
A current experimental
situation does not allow to resolve this effect because of the presence
of the normal component which introduces 
a substantial damping on its own. 

 Damping of the normal modes at finite $T$
of the confined condensate has been experimentally studied by the JILA
\cite{DAMP1} and the MIT \cite{DAMP2} groups. 
Recently 
it was stressed by Pitaevskii \cite{PITPRIV}  
(see also in Refs.\cite{PIT3,GIO,TOSI,VU})
that the damping in the trap containing condensate is essentially
Landau damping (LD) which is collisionless in nature. 
 In other words, it is rather the dephasing 
of the collective modes than their irreversible dissipation
\cite{KIN}. It is worth noting that 
the reversible nature of the LD in the classical
uniform plasma can be revealed
in the plasma echo effect \cite{KIN}.

Theoretical approaches \cite{PIT3,GIO,TOSI,VU}
 are based on applying the standard results of the perturbation
theory developed for an infinite and uniform medium. Accordingly,
a discrete structure of the quasiparticle spectrum is tacitly
replaced by an effective continuum \cite{PIT3,GIO,TOSI,VU}. 
In the recent work \cite{SHLAP} it has been emphasized 
that the existence of the LD is directly related to
a presence of the randomness in the spectrum of the anisotropic traps.
Conversely, in the isotropic trap the LD should be
suppressed because the quasiparticle spectrum is regular \cite{SHLAP}.
 
In this paper, we suggest that in a confined condensate where
the LD is suppressed 
 the damping of the collective 
modes can still be observed.  
Specifically, thermal fluctuations of the population
 factors of the normal
component induce a reversible dephasing of the collective 
oscillations. 
 The nature of such a damping  is similar to that of the
 dephasing observed in quantum dots (see in Ref.\cite{DOT}). 
We show that, while being essentially zero for a traditional
uniform condensate,
the rate of such a  dephasing $1/\tau_d$ in the trap
containing  $10^3 - 10^4$ atoms
can be comparable to the rate of the damping observed
experimentally in Ref.\cite{DAMP1}. 

We also suggest that
 reversibility of the damping in a confined condensate
can be tested in a kind of the phonon echo experiment
(see in Ref. \cite{MASON}),
when two consecutive external pulses imposed on the 
trap induce a third pulse -- the echo -- at the time 
approximately equal to  
twice the  time interval between the first two pulses.
In this paper we analyze the case of the isotropic trap,
where the LD is not expected to be an important mechanism 
of dissipation. 
In a future work we will investigate
the echo in strongly anisotropic traps, where the main mechanism of 
damping is the LD.

\section{Thermal fluctuations and dephasing in the atomic 
trap }

In the Refs. \cite{KUK,PIT2} it has been shown that a
collective mode of the confined
condensate should exhibit a 
dephasing caused by the inter-particle interaction. 
 In the case $T=0$,
the dephasing is produced by the particles forming
the mode. Below we will consider a similar effect caused by
the interaction between a low energy collective mode and the 
quasiparticles forming a normal cloud. 

Let us discuss general
reasons for the thermally induced dephasing 
in the presence of the inter-particle interaction.
Especially we will clarify why this effect is not significant
for an infinite condensate, where the only cause of the dephasing
should be  the LD. The following analysis is based on the approach
suggested by Pitaevskii in Ref.\cite{PIT2} for the case $T=0$.
We extend this analysis for $T\neq 0$. 

If an external modulation at some frequency $\omega_0$ has excited
a system, the many body time dependent wave function
$|t\rangle$ constructed
in terms of the quasiparticle states $|N_0,N_1 ...\rangle$ acquires 
the form

\begin{equation}
\displaystyle 
|t\rangle = \sum_{N_0,N_1 ...} C_{N_0,N_1 ...}{\rm e}^{-iE_{N_0,N_1 ...}t}
|N_0,N_1,...\rangle , 
\end{equation}
\noindent
where $N_n$ is a population number for the $n$-th 
($n=0, 1, 2, ...$) quasiparticle
state; $E_{N_0,N_1 ...}$ stands for the energy
of the state $|N_0,N_1 ...\rangle$
 and $ C_{N_0,N_1 ...}$ denote the  normalized coefficients 
($\sum_{N_0,N_1 ...} |C_{N_0,N_1 ...}|^2=1$). 
For the case of weak interaction between quasiparticles, the external
drive, which is in resonance with the energy of, e.g., the $0$-th quasiparticle
state, creates a coherent mixture of the quasiparticles 
in the $0$-th state and does not
affect significantly the other states. Accordingly, one can factorize
$ C_{N_0,N_1 ...}$ as

\begin{equation}
 C_{N_0,N_1 ...}=C_{N_0}^{(0)}
C^{(1)}_{N_1}C^{(2)}_{N_2}...   , 
\end{equation}
\noindent
and assume that \cite{PIT2} 

\begin{equation}
|C_{N_0}^{(0)}|^2= \frac{\bar{N}_0^{N_0}}{N_0!}\exp (-\bar{N}_0), 
\end{equation}
\noindent
with the rest of the $C$- coefficients corresponding to the
number states ($ C^{(n)}_{N_n} = 1$ for some $N_n=N'_n$ and
 $ C^{(n)}_{N_n} = 0$
otherwise). In Eq.(3) $\bar{N}_0$ is given by the value of the classical
amplitude of the resonant mode \cite{KUK,PIT2}. 

The expectation value of the single particle operator $A(t)$,
which changes the number of quasiparticles in the given state by 1,
acquires a resonant contribution
from the $0$-th state. This is 

\begin{equation}\begin{array}{r}
\langle t|A|t \rangle = \sum_{N_0} C^{(0)* }_{N_0+1}  C^{(0)}_{N_0}\bar{A}
\exp [i(E_{N_0+1, N_1, ...} - \\ \\
 E_{N_0, N_1, ...})t]
 + c.c. ,
\end{array}\end{equation}
\noindent
where $\bar{A}$ is a corresponding single-particle matrix element.  
In the case of a weak interaction between quasiparticles, one can
expand the energy in accordance with the Landau theory of quantum
liquids as

\begin{equation}
 E_{N_0, N_1, ...}= \sum_n \omega_nN_n + 
{1\over 2}\sum_{mn} g_{mn}N_mN_n + o(N_iN_jN_k),      
\end{equation}
\noindent
where the coefficients $g_{mn}$ arise due to interaction
between quasiparticles.  

At finite $T$, the solution (4) should be averaged over the initial
values of $N_n$ with $n>0$, distributed in accordance
with a thermal ensemble. Substituting (5) into (4) and 
neglecting the term $\sim N_0^2$ significant for very small $T$ only
(see \cite{KUK,PIT2}), one obtains 

\begin{equation}\begin{array}{l}
\displaystyle \langle t|A|t \rangle_T =\bar{A}{\rm e}^{i\omega_0 t}
 \sum_{N_0} C^{(0)* }_{N_0+1}  C^{(0)}_{N_0} \\  \\
\displaystyle \langle\exp (it\sum_{n>0} g_{0n} N_n) \rangle_T
 + c.c., 
\end{array}\end{equation}
\noindent
with $\langle ... \rangle_T$ denoting the thermal average with respect
to the population factors $N_n$. 

In what follows we will assume that $N_n$ are distributed in
accordance with the grand canonical distribution. 
As will be discussed below, this assumption is valid 
as long as the mean number of atoms in the condensate
$N_c$ is sufficiently large. Thus, 
after straightforward calculations one finds

\begin{equation}\begin{array}{l}
\displaystyle\langle\exp (it\sum_{n>0} g_{0n} N_n) \rangle_T= \\ \\
\displaystyle\exp\left( -\sum_{n>0} \ln \frac{1-\exp(ig_{0n}t-\omega_n/T)}
{1-\exp(-\omega_n/T)}\right). 
\end{array}\end{equation}

We note that in a large system the matrix elements $g_{0n}$
are scaled as $g_{0n}\sim 1/V$ by the effective volume $V$ 
of the system. Accordingly, in the formal limit
 $V\to\infty$,
 one should expand
$\exp(ig_{0n}t)$ in Eq.(7) up to the first term $\sim 1/V$ only.
Then a smallness of $1/V$ will be compensated by the summation
$\sum_n \sim V$. This
results in 

\begin{equation}
\langle\exp (it\sum_{n>0} g_{0n} N_n) \rangle_T=
\exp( it\sum_{n>0} g_{0n} \bar{N}_n), 
\end{equation}
\noindent
where $\bar{N_n}$
denotes a Bose distribution 
of the non-interacting quasiparticles. As one can see, in the infinite
system fluctuations of the numbers of the quasiparticles do not
cause any dephasing. Instead, the frequency $\omega_0$ experiences
a thermal shift $\omega_0 \to \omega_0 + \sum_ng_{0n} \bar{N}_n$,
and the only cause
of the dephasing is the LD.   

In the case of a finite system one should keep higher order terms
in Eq.(7). As will be shown below, the dephasing rate in the
oscillator
trap contains a smallness $a/r_{trap}<<1$, where $a$ and 
 $r_{trap}$ stand for the scattering length
 $ \sim 10^{-7}-10^{-6}$cm and some typical scale $\sim 10^{-4}$cm 
associated with the trapping potential, respectively. Therefore,
  as long as $a/r_{trap} << 1$,
one can neglect the higher terms $o(a^2/r_{trap}^2)$.
 Such an approximation corresponds
to expanding $\exp (ig_{0n}t) \approx 1 + ig_{0n}t - g_{0n}^2t^2/2
+o(g^3_{0n})$.
This results in Eq.(7) being rewritten
as

\begin{equation}\begin{array}{r}
\displaystyle\langle\exp (it\sum_{n>0} g_{0n} N_n) \rangle_T= \\ \\
\displaystyle\exp\left( i\sum_{n>0} g_{0n}\bar{N}_nt-t^2/\tau_d^2 \right ), 
\end{array}\end{equation}
\noindent
where the notation 

\begin{equation}
\displaystyle {1\over \tau_d}= \sqrt{{1\over 2}
\sum_{n>0} g^2_{0n}\bar{N}_n(1+\bar{N}_n)}
\end{equation}
\noindent 
for the dephasing rate is introduced. 

Eqs. (9) and (10) indicate that a collective mode of the
confined atomic condensate should exhibit
a dephasing induced by thermal fluctuations of the normal
component. 
Below we suggest a model of dephasing of a collective mode
of the confined condensate in the Thomas Fermi limit, and 
 calculate the value of $\tau_d$ for the 
isotropic trap. 

\section{Adiabatic effective action for the low energy 
collective modes}

In order to obtain $1/\tau_d$, one should find
the coefficients $g_{0n}$ and perform the summation in 
Eq.(10). An explicit expression for $g_{0n}$ can be 
derived from a many body 
 Hamiltonian $H$ taken
in the standard form 

\begin{equation}\begin{array}{l}
 H=\int d{\bf r}  \Psi^{\dagger}(H_1 - \mu) \Psi + H_{int}, \\
\displaystyle H_1= - {\hbar^2 \over 2m}\Delta + \sum_{i=1,2,3}
 {1\over 2} m\omega_i^2r^2 , \\
H_{int}={1 \over 2}u_o\int d{\bf r}\Psi^{\dagger}\Psi^{\dagger}\Psi\Psi, 
\end{array}\end{equation}
\noindent
where the Bose operators $\Psi , \Psi^{\dagger}$
obey the usual Bose commutation rule;
 $\mu$ is the
 chemical potential; $\omega_i$ denote three
frequencies characterizing the trapping potential;
and the interaction constant $u_o=4\pi \hbar^2 a/m$ is
expressed
in terms of the scattering length $a$ and the atomic mass $m$.
As usual, in the presence of the condensate, one uses
the conventional representation 

\begin{equation}
\Psi = \Phi_c + \Psi', 
\end{equation}
\noindent
where
$\Phi_c$ is a classical condensate wave function
giving the condensate density $n_c=|\Phi_c|^2$ 
and normalized to the number $N_c$ of atoms in 
the condensate; $ \Psi'$ stands for the excitation
part. For the latter, the Bogolubov
representation should be employed as 

\begin{equation}
\Psi'=\sum_n (U_na_n + V_n^*a^{\dagger}_n), 
\end{equation} 
\noindent
where $a_n$ destroys a quasiparticle on the level
having the energy $E_n$, and ($U_n, V_n$) is an 
eigenvector of the linearized Bogolubov equations

\begin{equation}\begin{array}{l}
E_nU_n=H'_1U_n + u_0\Phi_c^2V_n, \\ \\ 
-E_nV_n=H'_1V_n + u_0\Phi_c^{2*}U_n, \\ \\
H'_1=H_1 + 2u_0|\Phi_c|^2 -\mu.
\end{array}\end{equation}
\noindent
 Expressed in terms of
the quasiparticle operators $a_n, a_n^{\dagger}$, the
Hamiltonian (11) acquires the form

\begin{equation}
H= \sum_n E_n a^{\dagger}_n a_n +H_{int}, 
\end{equation}
\noindent
where the interaction part $H_{int}$ contains the terms 

\begin{equation}\begin{array}{c}
\displaystyle H'_{int}=\sum_{mnk} (g_{mnk} a^{\dagger}_m
 a_n a_k + H.c.) +\\   \\
\displaystyle \sum_{mnkl} g_{mnkl} a^{\dagger}_m a^{\dagger}_n 
 a_k a_l, 
\end{array}\end{equation}
\noindent
which could be identified with the terms $\sim N_mN_n$ in 
Eq.(5). 
Specifically,  performing calculations in the first
and in the second orders of the perturbation theory with respect
to $H'_{int}$, one finds
 the coefficients $g_{kn}$ in Eq.(5) as

\begin{equation}\begin{array}{c}
\displaystyle g_{kn}=4g_{knkn} + 2\sum_m 
|g_{mnk}|^2[\frac{2(E_n - E_m)} 
{(E_n - E_m)^2 - E_k^2} -\\ \\
\displaystyle {1\over E_n
+E_m - E_k} ] ,
\end{array}\end{equation}
\noindent
where the coefficients $g_{knkn}$ and $g_{mnl}$
can be expressed  explicitly in terms of the Bogolubov amplitudes 
 in Eq.(13) by means of substituting Eq.(12) in Eq.(11) and
selecting the terms (16). This yields 

\begin{equation}\begin{array}{r}
g_{knkn}={u_0\over 2}\int d{\bf r}[(|U_k|^2 + |V_k|^2)(| U_n|^2 + |V_n|^2)\\ \\
+ (U^*_kV_kV^*_nU_n + c.c)]
\end{array}\end{equation}
\noindent
and

\begin{equation}\begin{array}{r}
g_{mnk}=u_0\int d{\bf r}\Phi_c[U_k (U^*_mU_n + V^*_mV_n
+ U^*_mV_n  ) +\\  \\
 V_k(U^*_mU_n + V^*_mV_n+
 V^*_mU_n )]
\end{array}\end{equation}
\noindent
for real $\Phi_c$.
  
We note that  in the case of a continuum or quasi-continuum spectrum
\cite{SHLAP}, the sum in Eq.(17) gives the main contribution
to the imaginary part corresponding to the LD in the lowest order
of the perturbation theory \cite{PIT3,GIO,TOSI,SHLAP}.
In the isotropic trap, this imaginary contribution is not significant
\cite{SHLAP}, which implies that the LD is suppressed. Therefore, the real
value of $g_{kn}$ (17) could be employed in Eq.(10) for calculating
$1/\tau_d$.
Unfortunately, such an approach leads to an expression  
$1/\tau_d$ which formally diverges on low energies, despite the
natural expectation that the main contribution to the dephasing 
is produced by high energy excitations. Accordingly, the rate
(10) acquires the incorrect $T$-dependence $1/\tau_d \sim \sqrt{T}$
for $T> \mu$. In fact,  
this divergence can be eliminated by a proper renormalization
of the vertex in the low energy region, where an adequate
description relies on the hydrodynamical approach \cite{HYDRO}.
In order to solve this problem for the low
energy collective modes, we employ a simple scaling
procedure which yields a description of the effect of thermal dephasing
in closed form at finite $T$ in the limit of large $N$.

First, we note that the dephasing effect is an adiabatic process
when the high energy component follows the evolution of the low
energy collective mode without dissipating energy. As was discussed
in Ref.\cite{KAGAN,CASTIN}, the low energy confined condensate modes at $T=0$
can be viewed as 
time dependent scaling
$r_i \to r_i/b_i$ of the coordinates $r_i$ 
by some time dependent scaling variables $b_i=b_i(t)$.
Furthermore, it has been shown \cite{KAGAN} that,
if one ignores the kinetic energy term, such a scaling
approach is exact for any given initial state of the 
many body wave function. As will be seen below,
this implies that no dephasing of the low energy
scaling modes should occur in such an approximation.
The dephasing is induced by the kinetic energy terms,
which are, however, small in the Thomas Fermi limit
\cite{KAGAN}. Such a smallness implies that one can still
consider the scaling variables $b_i$ as proper collective
degrees of freedom, whose dynamics is modified in the presence
of the kinetic terms. Below we will derive
an adiabatic classical action for the $b$-variables.
In order to do this, we treat the $\Psi $-operator as a
classical field by means of considering the
$a,\,a^{\dagger}$ operators in Eq.(13) as $c$-numbers.
Then we employ the scaling ansatz  
\cite{KAGAN} 

\begin{equation}
\displaystyle
\Psi (r_i,t)\to {{\rm e}^{i\varphi}\over \sqrt{b_1b_2b_3}}
\Psi({r_i\over b_i},t'(t)),\,\, 
\displaystyle 
\varphi = {m\over 2\hbar}\sum_i{\dot{b}_i\over b_i}r^2_i,
\end{equation}
\noindent
where $t'(t)$ is some function of time determined in terms
of the variables $b_i$ and their time derivatives $\dot{b}_i$.
In what follows we assume that the scale invariant shape
of $\Psi $ is given by Eqs.(12)-(14) obtained
for $b_i=1$. 
 In this manner we eliminate the non-adiabatic processes
induced due to $\dot{b}_i\neq 0$. Consequently, one can
derive an effective classical action 
$S_b= S[b_i, \dot{b}_i]$ for the variables
$b_i$ by means of performing
the scaling transformation (20) in the full classical action

\begin{equation}
S=\int dt\{\int d{\bf r}
 {i\over 2}(\Psi^*\dot{\Psi} - h.c.) - H\}.
\end{equation}
\noindent 
Note that a substitution of the form (12), (13) is to be done
in Eq.(11) and, then, in Eq.(21).
Then,  
the off-diagonal products $a_n^*a_m$ 
 and $a_na_m$ ($m\neq n$) should be eliminated
in the adiabatic approximation because
these oscillate in time. 
The diagonal terms $a^*_na_n$, which do
not oscillate, should be retained and identified with the
population factors.
Let us denote the action obtained as a 
result of such a procedure as  $S_b=\langle S \rangle$.
Then, we find

\begin{equation}
S_b=\int dt \{\sum_i [{m\over 2}R_i(\dot{b}^2_i -
\omega_i^2b_i^2) - {P_i\over 2b^2_i}] - {G\over b_1b_2b_3}\},
\end{equation}
\noindent
where the notations

\begin{equation}\begin{array}{c}
\displaystyle R_i=\langle  \int d{\bf r}\Psi^*
 r^2_i \Psi  \rangle, \quad 
P_i= \langle \int d{\bf r}{\hbar^2 \over m}
\nabla_i\Psi^*
\nabla_i \Psi  \rangle, \\
\displaystyle G= \langle {u_o\over 2}
 \int d{\bf r}\Psi^*\Psi^*
\Psi \Psi \rangle
\end{array}\end{equation}
\noindent
are introduced. Taking into account that these
 quantities do not depend
on $b_i$, one can  
 vary $S_b$ with respect to $b_i$ and obtain
the classical equations of motion

\begin{equation}
\ddot{b}_i + \omega_i^2b_i - {G\over mR_i} 
{1\over b_ib_1b_2b_3}
-{P_i\over mR_i} {1\over b^3_i}=0,\,\, i=1,2,3.
\end{equation}

We note that these equations reproduce correctly
the $T=0$ low energy spectrum obtained in Ref.\cite{STRING}
for the isotropic trap characterized by the relation
$\omega_1=\omega_2=\omega_3=\omega_{ho}$.
Indeed, in the case of zero temperature one should take
the means (23) over the condensate state
by setting $a_n=a^*_n=0$ in Eq.(13), and, accordingly,
taking $\Psi = \Phi_c$ in Eq.(23). Then, the virial
theorem yields 

\begin{equation}
m\omega^2_iR^{(c)}_i - P^{(c)}_i-  G^{(c)} =0,
\end{equation}
\noindent
with the superscript ${(c)}$ indicating that the means in Eq.(23)
are taken over the condensate state. Employing this relation
in Eqs.(24), one finds

\begin{equation}
\ddot{b}_i + \omega_i^2b_i - \omega_{i}^2(1 -
\xi^{(c)}_i ) {1\over b_ib_1b_2b_3}
-\xi^{(c)}_i\omega_{i}^2 {1\over b^3_i}=0,
\end{equation}
\noindent
where
 the notation

\begin{equation}
\xi^{(c)}_i= {P^{(c)}_i\over m\omega_{i}^2R^{(c)}_i}
\end{equation}
\noindent
is introduced. We note that $\xi^{(c)}_i$ determines
the ratio of the averages of the kinetic energy to the 
harmonic potential energy both taken for the $i$-direction.
Accordingly, for the case of the isotropic trap considered
in Ref.\cite{STRING} one can find $\xi^{(c)}_i=
E_{kin}/E_{ho}$, where $E_{kin}$, $E_{ho}$ stand
for the total kinetic and the total
 harmonic energies, respectively.
Then, from the linearized version of Eqs.(26) 
we reproduce the frequencies
of the quadrupolar mode $\omega_Q = \sqrt{2}\omega_{ho}
(1+ E_{kin}/E_{ho})^{1/2}$
and of the breathing mode $\omega_M = \omega_{ho}
(5- E_{kin}/E_{ho})^{1/2}$ derived in Ref.\cite{STRING}.
 
Note also that for $T\neq 0$, 
 Eqs.(24)  coincide with those
derived in Refs.\cite{KAGAN,CASTIN} in the Thomas-Fermi limit.
 This can be seen by  
means of setting the kinetic energy terms $P_i=0$ in Eqs.(24) and
performing the scaling transformation

\begin{equation}
b_i= \kappa_i\tilde{b}_i 
\end{equation}
\noindent
with some $\kappa_i$ chosen in such a way as to
make the solution $\tilde{b}_i=1$ to be the
equilibrium one.           
Furthermore, it can be seen that  
in the isotropic 2D case,
  when only two scaling
variables $b_1,\,b_2$ should be taken into account,
the dependence on $P_i$ can be eliminated 
by the scaling transformation (28), so
that
the frequency of the breathing mode does not 
depend on $P_i$. This implies that
no dephasing of the 2D breathing mode 
should occur in accordance 
with the result of exact calculations
of Ref.\cite{2D}. 
 
In order to simplify the following analysis of the $T\neq 0$ case,  
 let us consider a breathing mode in the 3D isotropic trap.
Thus, we set $b=b_1=b_2=b_3$ and $b=\kappa \tilde{b}$
 in the transformation (20)
as well as in the action $S_b$ (21).
 Then, varying $\delta S_b/\delta b=0$, 
we obtain the nonlinear 
equation describing the low energy adiabatic dynamics 
of the breathing mode in the 
presence of the normal component: 

\begin{equation}
\ddot{\tilde{b}} + \omega_{ho}^2\tilde{b}
 - {\omega_{ho}^2\over \tilde{b}^4} 
+ \xi \omega_{ho}^2 {1-\tilde{b} \over \tilde{b}^4}=0, \quad 
\displaystyle \xi = {P \over mR\omega^2_{ho}\kappa^4},
\end{equation}
\noindent
where 
$P=P_1+P_2+P_3$, $R=R_1+R_2+R_3$
and $\kappa$ obeys the relation 

\begin{equation}
\omega^2_{ho} - {P \over m\kappa^4R}
-{3G\over m\kappa^5R}=0.
\end{equation} 
\noindent
An explicit form of $\xi$ can be found in the limit $P\to 0$,
which
corresponds to neglecting the kinetic energy of the system
if compared with the interaction energy.
Setting $P=0$ in Eq.(30), one finds $\kappa$ and then,
resorting back to Eq.(29), obtains

\begin{equation}
\xi = {P\over \omega^{2/5}_{ho}(mR)^{1/5}(3G)^{4/5}}.
\end{equation}
\noindent
In order to express $P,\,R,\,G$ in terms of the
products $a^*_na_n$ which, as was mentioned above,
 should be then identified
with the  population
factors $N_n=a^*_na_n$ of the quasiparticles
in the second quantized picture,
 one should use the representations (12), (13), (23).
However, in the Thomas Fermi limit valid for large numbers
$N_c$ of atoms in the condensate, one may neglect
corrections to $R$, $G$ and $N_c$ due to the excitations at 
not very large temperatures. Indeed,
in the condensate state $R\sim G \sim N_cr^2_c$,
where $r_c$ stands for the condensate radius 
 $r_c \sim N_c^{1/5}$ \cite{BAYM}.
Therefore, $R\sim G \sim N_c^{7/5}$. The kinetic
term $P\sim N_c/r_c^2 \sim N^{3/5}_c$. High energy excitations
produce changes $\delta R,\, \delta G,\,\delta N_c,\,\delta P$ 
of $R,\, G,\, N_c,\, P$, respectively. 
Therefore, their relative contributions $\delta R /R$,
$ \delta G/G$, $\delta N_c/N_c$ and $ \delta P/P$ to the value of
$\xi $ are very different. Specifically, the first two terms
contain an additional smallness $\sim N^{-4/5}_c$ if compared
with the last one. Therefore, in calculating $\xi $ in Eq.(31),
one should take into account only the contribution due to $\delta P$,
and take the values $R,\, G$ determined for the condensate state
by Eq.(23) for some mean value of $N_c$.

The term $\delta N_c/N_c$, arising due to the conservation of 
the total number
of particles, can be neglected as well. 
Indeed, if the normal component fluctuates by
some number $\delta N'$, the conservation
of the total number implies that $\delta N_c =-\delta N'$.
The kinetic term fluctuates as $\delta P
\sim \delta N'$. Therefore, the ratio $\delta N_c/N_c $ over 
$\delta P/P$ contains a smallness $\sim N_c^{-2/5}<<1$.

  Finally, employing representations
(12), (13) in (23) and taking the means so that
$\langle a^{\dagger}_ma_n \rangle = N_n\delta_{mn}$, one finds 

\begin{equation}\begin{array}{l}
\xi =\xi^{(c)} + \sum_n g_nN_n,  \\  \\
\displaystyle g_n=-{\hbar^2\over m^2R_c\omega^2_{ho}}\int d{\bf r}
(U^*_n\Delta U_n + V^*_n\Delta U_n), \\  \\
R_c=\int d{\bf r}r^2|\Phi_c|^2,
\end{array}\end{equation}
\noindent
where the virial relation (25) has been utilized, and
$\xi^{(c)}$ denotes the contribution due to the kinetic
energy of the condensate. In
what follows we will neglect this term which 
does not depend on $N_n$.

The solution $\tilde{b}$ of Eq.(29) should be averaged
over $\xi$ represented by Eq.(32). Such an averaging can be understood
as a thermal ensemble averaging over possible Fock states of
the quasiparticles. This interpretation closely resembles
the case of destructive measurements 
\cite{DAMP1}, when the initial conditions determined by  $N_n$
 for each newly
created condensate can vary from one condensate
to another.
In the case of non-destructive measurements, 
the averaging of the solution $\tilde{b}$ should be performed
over a single many body state, which is a mixed
state rather than a pure Fock state with respect to $N_n$. In accordance
with the ergodic hypothesis such an averaging
should give the same result as that obtained by means of 
thermal averaging as long as the 
number of quasiparticles is sufficiently
large. In what follows we will not distinguish these
two cases, and will employ the thermal averaging
 $\langle ... \rangle_T$ with respect to $N_n$.
This averaging can be performed explicitly for the
linearized solution of Eq.(29). Specifically, representing
 $\tilde{b}=1+ \eta $ in Eq.(29) and keeping the terms linear 
in $\eta$, one finds 

\begin{equation}\begin{array}{l}
\displaystyle 
\langle \eta (t) \rangle_T = \eta_0\langle {\rm e}^{i\omega_{ho}(
5 - \xi)^{1/2}t}\rangle_T + c.c. \sim \\  \\
\displaystyle {\rm e}^{i\bar{\omega}_Mt 
- t^2/\tau^2_{dM}} + c.c., \\ \\
\displaystyle \bar{\omega}_M =\sqrt{5}\omega_{ho}
-{1\over 2\sqrt{5}}\langle \xi \rangle_T.
\end{array}\end{equation}
\noindent
where $\eta_0=const $ accounts for the initial condition
$\tilde{b}(0)=1+\eta_0$, and the dephasing rate of the 
breathing mode is

\begin{equation}\begin{array}{l}
\displaystyle {1\over \tau_{dM}}=\omega_{ho} \sqrt{{1\over 40}
\sum_n g^2_n\bar{N}_n(1+\bar{N}_n)},\\ \\ 
\displaystyle\bar{N}_n=
{1\over \exp ({E_n\over T})-1},
\end{array}\end{equation}
\noindent
 with the coefficients $g_n$ given by Eq.(32).
Performing similar calculations
for the quadrupolar mode, we find the relation $1/\tau_{dQ}
=\sqrt{5/2}/\tau_{dM}$ for the dephasing rate of the
quadrupolar mode. 
Taking into account Eqs.(32), (34), we obtain an explicit
expression for the rate $1/\tau_{dM}$ (34) in the WKB approximation
\cite{WKB} (see Appendix A) as

\begin{equation}
{1 \over \tau_{dM}}=\Gamma_M D(T/2\mu),
\end{equation}
\noindent
where the coefficient $\Gamma_M$ is

\begin{equation}
\Gamma_M= {35 \over \sqrt{5\pi}}\left({r_{ho}\over r_c}\right)^2
{a\over r_{ho}}\omega_{ho}, \,\, r_{ho}=\sqrt{{\hbar \over m\omega_{ho}}},
\end{equation}
\noindent
and the universal dimensionless function $D(\beta )$ is defined in Appendix A
(see  Fig.1), 
with the parameters $r_c,\,\, \mu$ given explicitly in (A2). 

In the limits $\beta >> 1$ ($T>> 2\mu$) and $\beta <<1$ ($T<< 2\mu$) 
the function $D(\beta)$ can be found explicitly 
(see Eqs. (A28) and (A32), respectively). The
current experimental situation is closer to the first case.
It is convenient to express $T$ in units of the transition
temperature $T_c$ of the Bose-Einstein condensation
in the isotropic oscillator trap 

\begin{equation}
T_c=\hbar \omega_{ho}\left({N \over \zeta (3)}\right)^{1/3},
\end{equation}
\noindent
where $\zeta (3)\approx 1.202$; $N$ is the total number of the trapped atoms
(for $T$ not very close to $T_c$ we set $N_c\approx N$). Then, we find 

\begin{equation}\begin{array}{l}
\displaystyle{1\over \tau_{dM}}={35\sqrt{0.3}\over 15^{7/5}(\zeta (3))^{5/6}}
\left({T\over T_c}\right)^{5/2}\left({r_{ho}\over a}\right)^{2/5}
N^{-17/30}\omega_{ho}
\end{array}\end{equation}
\noindent
from Eqs.(35)-(37), (A28), (A29).
Choosing the values $T/T_c =0.9,\, N=2\cdot 10^3-10^4$
 and $\omega_{ho} = 2\pi
\cdot 200 s^{-1},
\, r_{ho} =10^{-4}$cm typical for the experiment \cite{DAMP1}, we
obtain the rate $1/\tau_{dM} \approx 40 s^{-1} - 20 s^{-1}$. 
We note that these values of $1/\tau_{dM}$ are close to the
 damping rate observed in Ref.\cite{DAMP1}. However, 
for the chosen parameters, 
$\beta =T/2\mu \approx 1.3$ which is far from the requirement 
$\beta >>1$ insuring the validity of Eq.(38). 
Nevertheless, the above estimates remain valid. 
Indeed, evaluating the complete expressions (A18), (A19) numerically
(see Fig.1) changes
these estimates by only about 20\%. Specifically, the rate
becomes $\approx 50 s^{-1} - 25 s^{-1}$. 
For the lowest temperature achieved in the experiment \cite{DAMP1}
$T\approx 0.4 T_c$, Eq.(38) becomes invalid because this
temperature corresponds to $\beta \approx 0.6<1$. 
Accordingly, a numerical evaluation of $D(\beta)$ by means of 
Eqs. (A18), (A19) 
and, then, a substitution of the result into Eq.(35) yields the rate
$1/\tau_{dM} = 8 s^{-1} - 4 s^{-1}$ which is also in the range 
obtained by the JILA group \cite{DAMP1}.

\begin{figure}[hbt]
\epsfxsize=\columnwidth\epsfbox{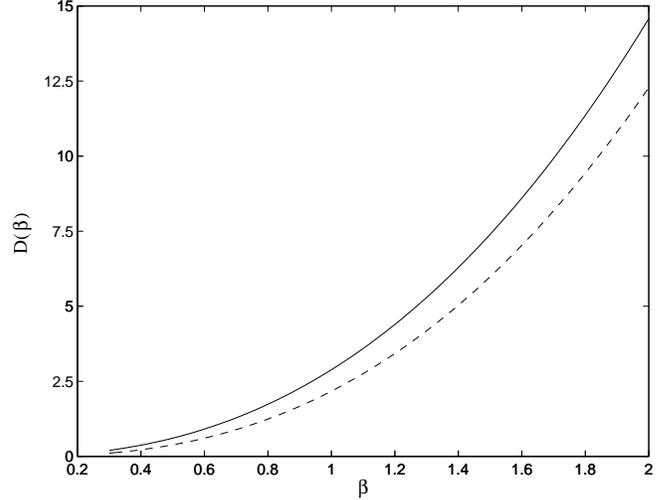}
\caption{The universal function $D(\beta)$: the solid
line is the result of numerical calculation (A18),(A19); the dashed line is
the approximation (A28).}
\label{fig.1}
\end{figure}
We emphasize
that in the anisotropic trap employed by the JILA group the damping
is most likely to be caused by 
the LD  
\cite{PITPRIV,PIT3,GIO,TOSI,VU,SHLAP}, and not by the mechanism
discussed above. The discussed mechanism in its pure form can be realised
in the isotropic trap only.  Therefore, a correspondence between
the rate calculated above
for the isotropic trap and that measured by the JILA group
\cite{DAMP1} for the anisotropic trap
 indicates that, while decreasing a degree of the
trap anisotropy, the damping rate should practically stay unchanged
despite the fact that the nature of the damping changes.

In the case $T<< 2\mu$ one could use an explicit form (A32) for
$D(\beta)$ and, correspondingly, find an explicit $T$-dependence
of the rate (35). However, in this case the rate becomes so small,
that the mechanism of the quantum self-dephasing \cite{KUK,PIT2}
comes into play.

It is interesting to investigate the dependence of the 
dephasing rate on the amplitude of the oscillations. It is
worth noting that in the case $T=0$ such a dependence 
is very pronounced \cite{KUK,PIT2}. 
In contrast, as will be seen below, the amplitude dependence 
at finite $T$ is weak. Indeed,
this dependence is due to
the nonlinearity of the term $\sim \xi$ in Eq.(29).
In the lowest order with respect to the initial value  
$\eta_0$ in Eq.(33), this dependence
can be obtained by expanding Eq.(29) up to the terms
$\sim \eta^2$ and $\eta^3$ and finding the correction
to the frequency of the lowest harmonic in the order
$\sim |\eta_0|^2$. Performing straightforward calulations
(see Appendix B) and then averaging over the ensemble, we obtain

\begin{equation}
{\tau_{dM}\over \tau_{dM}(A_1)}=1- {7\over 3} |A_1|^2
\end{equation}
\noindent
the ratio of the rate $1/\tau_{dM}(A_1)$ determined
in  the second order with respect to the amplitude $A_1=2\eta_0$ of
the collective mode 
to the rate $1/\tau_{dM}$ in the 0-th order given by Eqs.(34), (35). 
As one can see, the rate demonstrates a slow decrease 
as a function of the amplitude $A_1<<1$. 

Here we have shown that 
the collective excitations of the confined
Bose-Einstein condensate should demonstrate a dephasing  
caused by thermal fluctuations of the normal 
component. 
In the following section we will discuss how
this dephasing effect can be distinguished from
irreversible dissipation experimentally.

\section{The Echo effect in a confined Bose-Einstein
condensate}

The reversible nature of the damping can be tested
in an echo experiment similar to 
the spin echo, the photon echo and the phonon echo 
effects (see in Ref.\cite{MASON}).    
The nature of this effect can be briefly outlined
as follows \cite{MASON}. A short external pulse imposed on the
system at the time $t=0$ excites a collective
mode. The collective mode amplitude 
 decays due to dephasing as well as
due to irreversible dissipation. Both processes
are characterized by their typical rates
$1/\tau_d$ and $\gamma$,
 respectively. The second pulse imposed
at the time $t=\tau$  
partly reverses in time
the evolution of the system initiated by the first pulse.
 This implies a partial revival
of the dephased amplitude at the time $t \approx 2\tau$.
We note that the occurrence of the echo is a general property
of the system where irreversible damping is weaker
than the dephasing. Thus, 
a necessary condition for observing a distinct
echo is
$\tau_d < 1/\gamma, \quad \tau_d < \tau < 1/\gamma$ .

Specific features of the echo depend on the details 
of the system. The time profiles
of the responses, as Eq.(33) indicates, 
 should be gaussian
 in the case of the thermal dephasing
discussed above. In the case of the LD 
these responses should be characterized by 
exponential relaxation. Presently
available experimental data \cite{DAMP1,DAMP2}
do not allow the distinguishing of the gaussian type
damping from the exponential one \cite{PRCORN}.
 In the next paper we will analyze
the echo in the anisotropic confined condensate,
where the main cause of the damping is the LD. Below
we will study the situation in the isotropic trap, where the
dephasing is caused by the thermal mechanism described in Sec.III.

A relevant description for the case under consideration
 relies on Eq.(29) modified to incorporate the external
drive as well as some possible irreversible dissipation. 
As was discussed in Ref.\cite{KAGAN},
 the external drive $\delta\omega^2(t)$, which
changes the curvature of the trapping potential,
should be included in the linear part of the equation
for the scaling variable $\tilde{b}$. Accordingly, 
Eq.(29) is rewritten as

\begin{equation}
\ddot{\tilde{b}} + [\omega_{ho}^2 + \delta \omega^2 (t)] \tilde{b} - 
{\omega^2_{ho} \over \tilde{b}^4}
+ 2 \gamma \dot{\tilde{b}} +\xi \omega^2_{ho}{1- \tilde{b}\over
\tilde{b}^4}=0.
\end{equation}
\noindent
For  $\xi =\gamma =0$, one obtains the equation 
derived in Refs.\cite{KAGAN,CASTIN} for the case $T=0$.
The term $\sim \gamma$ describes the irreversible dissipation
at $T\neq 0$. The term $\sim \xi$ introduced already in Eq.(29),
with $\xi$ given by Eq.(32),
accounts for the dephasing effect discussed above. 
 
The time dependent part $\delta \omega^2(t)$
of the frequency  
should be driven so as to 
be in resonance with the collective
mode, that is in the form

\begin{equation}
\delta \omega^2(t) =  - f(t)\exp (i\omega_0 t) -
f^*(t)\exp (-i\omega_0 t),
\end{equation}
\noindent
where $\omega_0=\sqrt{5}\omega_{ho}$ and $f(t)$ stands for the complex amplitude of the external drive.
In order to avoid exciting other modes of the system,
this amplitude should be considered as a slow envelope of
the resonant drive with a typical time $\tau_f >> \omega^{-1}_0$.
The echo, then, can be produced by making $f(t)$ reach a maximum
at $t=0$ and then become zero untill the time $t=\tau$,
when $f(t)$ peaks again. For sake of simplicity, 
we will ignore 
other modes and will 
analyze the simplest situation when the external drive
produces two $\delta$-pulses 

\begin{equation}
\delta \omega^2(t)=- f_1\delta(t) - f_2\delta(t-\tau)
\end{equation}
\noindent
at $t=0$ and $t=\tau $ having amplitudes $f_1, f_2$, respectively. 

For the case of small amplitudes $f_1,\, f_2$ of the drive, one
should  
 look for an evolution of the
small perturbation around the equlibrium value $\tilde{b}=1$.
We note that, in contrast with the conventional
situation \cite{MASON}, the echo response
in our model does not require non-linearity
of the dynamical equation. This is due to the fact
that the external drive plays a two-fold role. 
Specifically, on one hand, it gives rise to an effective external 
force $- \delta \omega^2(t) $ and, on the other hand, it excites the system
parametrically. Indeed, linearizing Eq.(40) by 
the substitute $\tilde{b}= 1+ \eta$, with $\eta <<1$, one obtains

\begin{equation}
\ddot{\eta} + [\omega^2_{ho}(5 -\xi) + \delta \omega^2(t)]\eta
 + 2 \gamma\dot{\eta} = 
- \delta \omega^2(t),
\end{equation}
\noindent
where the higher order terms of $\eta$ are neglected. 

We assume that
initially at $t= -\infty$ the mode was not excited
($\eta (-\infty)=\dot{\eta}(-\infty)=0$). 
Then taking into account Eq.(42), one finds form Eq.(43)

\begin{equation}
\eta (0)=0,\quad \dot{\eta}(0)= f_1
\end{equation}
\noindent
after the first pulse. 
The second pulse at $t=\tau$ results in a jump of $\dot{\eta}$
so that 

 \begin{equation}\begin{array}{l}
\dot{\eta}(\tau + \varepsilon)=
\dot{\eta}(\tau - \varepsilon) +  f_2(1 + \eta(\tau)), \\
\eta (\tau + \varepsilon)= \eta (\tau - \varepsilon)=\eta (\tau),
\end{array}\end{equation}
\noindent
where $\varepsilon \to +0$. 

We are looking for a solution at $t > \tau $.
It has the form

\begin{equation}\begin{array}{l}
\eta (t)= A{\rm e}^{(iQ - \gamma )(t - \tau)} + 
A^*{\rm e}^{(-iQ - \gamma )(t - \tau)},\\
\displaystyle Q = \omega_0(5- \xi)^{1/2}\approx
\sqrt{5}\omega_0(1 - {\xi \over 10}), 
\end{array}\end{equation}
\noindent
where we have 
taken into account that $\gamma << \omega_{ho}$
and $\xi << \omega_{ho}$. 
An explicit expression for the coefficient $A$
can be obtained if one employs the  
conditions (44), (45). Finally, we find
the solution (46) for $ t > \tau$ expressed as

\begin{equation}\begin{array}{c}
\eta (t)= {f_1\over 2iQ}(1+ {f_2\over 2iQ})
{\rm e}^{(iQ - \gamma)t} +
 {f_2\over 2iQ}{\rm e}^{(iQ - \gamma)(t - \tau)} 
 +\\ \\ 
\eta _e(t) + c.c.
\end{array}\end{equation}
\noindent
where

\begin{equation}
\eta_e(t) = {f_2f_1 \over 4Q^2}{\rm e}^{iQ(t - 2\tau) - \gamma t}
 \end{equation}
\noindent
 represents the echo occurring at the time moment $t=2\tau$. 
Indeed, after thermal averaging over $N_n$,
one finds

\begin{equation}\begin{array}{r}
\displaystyle \langle \eta (t)\rangle_T = {f_1\over \omega_0}[\sin (\omega t)
- {f_2\over 2 \omega_0} \cos (\omega t)]
 {\rm e}^{- \gamma t - t^2/\tau_{dM}^2} +\\  \\
\displaystyle  + {f_2\over \omega_0}\sin (\omega (t - \tau)){\rm e}
^{- \gamma (t - \tau) - (t - \tau)^2/\tau_{dM}^2} \\  \\
\displaystyle +{f_2f_1\over 10\omega_{ho}^2}
\cos (\omega (t - 2\tau)){\rm e}^{ - \gamma t
- (t - 2\tau)^2/\tau_{dM}^2}.
\end{array}\end{equation}
\noindent
The echo amplitude can be defined as $A_e=\langle \eta (2\tau )\rangle_T$
in the case when
the last term in Eq.(49) dominates the sum. In this case we find

\begin{equation}
A_e={f_2f_1\over 10\omega_{ho}^2}{\rm e}^{-2\gamma \tau}.
\end{equation}
\noindent
In deriving Eqs.(49), (50) 
in the limit under consideration, we have made the replacement
$Q =\sqrt{5} \omega_{ho}$ everywhere in 
Eqs.(47), (48) except in the exponents,
where $Q$ linearized in $\xi$ and given by Eq.(46) has been 
employed. 
Thus, we obtained the echo effect
in the linear approximation.

We have also analyzed the non-linear echo problem
for Eq.(40) numerically. This equation
was solved for a given value of $\xi$, and then
the final solution was averaged over 
the values of $\xi$ distributed in accordance with
the gaussian $G(\xi) = \theta /\sqrt{\pi}\exp(-\xi^2/\theta^2 )$,
where $\theta$ determines the effective width of the distribution
in such a way that the averaging of the linearized solution
reproduces the result (33), (34). Specifically, we set 
$\theta = \sqrt{80}/ \omega_{ho}\tau_{dM}$.
 The results of the calculations
are shown in Fig.2.  
In the case a) the amplitude of the second pulse
is too small to make the echo observable. In the case b) the second 
amplitude is 5 times stronger so that the echo is distinct. In the case
c), while the second pulse amplitude 
$f_2$ is the same as in the cases a) and b),
 the amplitude of the first pulse $f_1$ is two times
larger than that in the cases a),b)
(note the different scale of the vertical axis in the case c)).
   As one can see, the echo in this case
merges with the tail of the second pulse, which creates an impression
that the decay time of the second pulse increases by several times. 
In order to produce the echo in the case of the large amplitudes  
$f_1,\, f_2$, the time separation $\tau$ between the pulses 
should be increased. However, in this case the irreversible dissipation
may strongly suppress the echo in accordance with Eq.(50).   

\begin{figure}[hbt]
\epsfxsize=\columnwidth\epsfbox{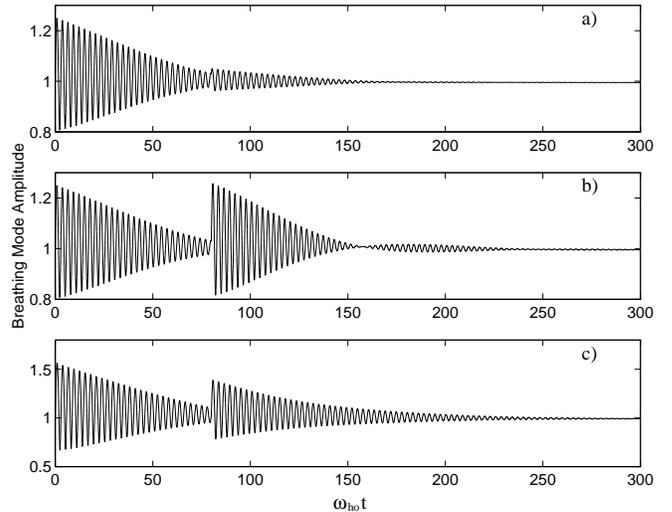}
\caption{The
echo effect in a confined Bose-Einstein condensate:
the  solution for the breathing mode $\tilde{b}(t)$ given
by Eqs.(40), (42)
( $\gamma  = 0.01\omega_{ho}$, 
$\theta^2 =0.02$,
 $\tau = 80\omega^{-1}_{ho}$).
The cases a), b) and c) are different in amplitudes $f_1,\,f_2$:
 a) $ f_1=0.5\omega_{ho},\,f_2=0.1\omega_{ho}$;
b)
$f_1=f_2=0.5\omega_{ho}$; c) $f_1=1\omega_{ho}, \, f_2=0.5\omega_{ho}$.
}
\label{fig.2}
\end{figure}
 
The echo effect analyzed above is a classical mechanics effect. Below certain
temperature $T_Q$, the rate of the quantum dephasing \cite{KUK,PIT2}
should become faster than the damping induced by the normal component.
Accordingly, the classical treatment employed above becomes no
longer valid. The problem should be reformulated in terms of the 
quantum dynamics of the variable $\eta $ in a sense of the approach
\cite{PIT2}, with the external drive (42) taken into account.
It can be shown that the echo
 still exists at $t=2\tau$. Therefore, the spontaneous quantum revival,
determined
by the interaction constant and thereby occuring at very long times 
\cite{PIT2}, can be induced to occur at
much shorter times comparable with the time of the quantum collapse
\cite{PIT2}. This problem will be considered in a separate publication. 

\section{Discussion }
We have suggested a mechanism for the apparent damping
of a Bose-Einstein condensate confined in the isotropic
oscillator trapping potential. This damping is a reversible
dephasing of the collective modes 
caused by thermal fluctuations of the population factors 
of the normal component. The calculation of the dephasing rate 
gives a value which is comparable with the experimentally
observed rate of the damping of the low energy collective modes 
in the atomic traps. 

This mechanism of dephasing 
relies on the ensemble averaging of the collective mode over the
initial population of the normal component. Thus, an assumption
is made that for any given initial distribution of the population
factors of the "hot" quasiparticles, this distribution does not
relax to equilibrium during the time of the dephasing $\tau_d$. 
Accordingly, processes of relaxation due to the LD or collisions
may  suppress the discussed mechanism, if their relaxation
times are comparable with  $\tau_d$. As long as the collisional 
damping introducing
irreversibility is unlikely to be relevant for such small temperatures
and densities, the LD is the only competing mechanism. However,
the LD is expected to be significant for substantially anisotropic traps
only \cite{SHLAP}.
Therefore, in the traps characterized by small anisotropy our mechanism 
should dominate.
  
Both mechanisms of damping --  Landau damping
and that considered above -- are reversible in nature, and therefore the 
evolution of the system can be partly reversed in time.
We suggest testing this  
in the atomic traps by employing the echo effect. As our analytical
and numerical calculations for the breathing mode in
the isotropic trap indicate, the echo amplitude as well as its
position depend on the parameters of the external drive which can be varied
over a wide range.  

\acknowledgements

The authors greatly appreciate very useful 
discussions with Lev Pitaevskii,
Eric Cornell and Joseph Birman.
 We also thank Alfred Levine and William Schreiber for stimulating
conversations related to this work. 

This work was supported by the PSC-CUNY Research Award Program.

\appendix
\section{WKB calculation of the dephasing rate}

The WKB calculation of the dephasing rate presented
here is essentially based on the results of Ref.\cite{WKB}.
Employing Eqs.(14) as well as the normalization condition
$\int d{\bf r}(|U_n|^2 - |V_n|^2)=1 $ in Eq.(32), one finds

\begin{equation}
\begin{array}{c}
\displaystyle g_n={2\over mR_c\omega^2_{ho}}\{E_n - \\  \\
\int d{\bf r}[({m\omega^2_{ho}r^2\over 2} + 2|K|^2 - \mu)
(|U_n|^2 + |V_n|^2)
 +\\  \\ (KV_nU^*_n 
 + c.c.)]\},
\end{array}
\end{equation}
\noindent
where the notation $K=u_o\Phi^2_c$ is introduced, and 
for the condensate wave function $\Phi_c=\sqrt{n_c}$
we employ the Thomas Fermi solution
\cite{BAYM}

\begin{equation}\begin{array}{l}
n_c={m\omega_{ho}^2 \over 2u_0}(r^2_c - r^2)\Theta(r_c - r), \\  \\
 r_c=r_{ho}\left({15N_ca \over r_{ho}}\right)^{1/5}, \,\,
 \mu={m\omega^2_{ho}r^2_c \over 2},
\end{array}\end{equation}
\noindent
 were $\Theta(z)$ is the step function; $u_o$ and $r_{ho}$
are defined in Eq.(11) and Eq.(36), respectively.
 Accordingly,
one finds the value of $R_c$ in Eq.(32)  as

\begin{equation}
R_c={r^7_c\over 35ar^4_{ho}}.
\end{equation}

States in the isotropic trap can be classified
in terms of the angular momentum $L$, its
$z$-component $L_z$ and the radial quantum number $n_r$.
Thus, the index in (A1) as well as in the sum (34) should
be understood as consisting of these three quantum numbers.
This implies that the summation $\sum_n ...$ in Eq.(34) runs over three 
quantum numbers
$n=(n_r,L,L_z)$. Because of the spherical symmetry,
 the summation over $L_z$
can be performed trivially, which gives
$\sum_n ...= \sum_{n_r}\sum_L(2L+1)...$. 
As will be seen below, the large values $n_r>>1,\, L>>1$
dominate this sum. Therefore, we replace the summation by 
the integration over $n_r,\, L$ 

\begin{equation}
\sum_n...\approx \int^{\infty}_0 dn_r \int^{L_0}_0dL 2L ...,
\end{equation}
\noindent 
where the upper limit $L_0$ is to be determined, and we made
the replacement $2L+1 \approx 2L$. 
It is convenient
to change the variable $n_r$ to $E$ by employing the quantization
condition \cite{WKB} 

\begin{equation}\begin{array}{l}
\displaystyle n_r+{1\over2}={1\over \pi \hbar}\int^{r_2}_{r_1}dr p_r,
\\ \\
p_r=\sqrt{2m(\sqrt{E^2 + |K|^2} - U_{eff}(r))},
\end{array}\end{equation}
\noindent
where

\begin{equation}
 U_{eff}(r)={1\over 2}m\omega^2_{ho}r^2 + {\hbar^2(L+1/2)^2\over 2mr^2}+
2|K| - \mu
\end{equation}
\noindent
denotes the effective WKB potential \cite{WKB},  and the turning
points $r_1,\, r_2$ obey the equation $p_r=0$ or

\begin{equation}
\sqrt{E^2 + K^2} - U_{eff}(r)=0.
\end{equation}
\noindent
Then the integral (A4) acquires the form

\begin{equation}
 {2\over \pi\hbar}
\int^{\infty}_0 dE \int^{L_0}_0dL L \int^{r_2}_{r_1}{dr\over v_r}...
\end{equation}
\noindent
where Eq.(A5) has been employed, and $v_r$ stands for the WKB
radial velocity \cite{WKB}. 

Before we proceed, it is convenient to employ dimensionless variables 
of length, energy and angular momentum as

\begin{equation}
x = {r \over r_c},\, \epsilon = {E\over \hbar \omega_{ho}} {r_{ho}^2\over
r_c^2},\, J =L {r^2_{ho}\over r^2_c},
\end{equation}
\noindent
respectively. Note that in these units the condensate radius $r_c$
equals 1, and the chemical potential $\mu$ and the quantity
$K$ become

\begin{equation}
 \mu'=  {\mu\over \hbar \omega_{ho}} {r_{ho}^2\over
r_c^2}={1\over 2}, \quad k ={1 \over 2}(1- x^2)\Theta(1-x),
\end{equation}
\noindent
respectively.
Accordingly, Eq.(A7) yields two sets of solutions for the dimensionless
turning points $x_{1,2}=r_{1,2}/r_c$

\begin{equation}
x_1=\sqrt{y_0},\quad x_2=\sqrt{b_+}, \quad J < \sqrt{2\epsilon}
\end{equation}
\noindent
where

\begin{equation}\begin{array}{l}
\displaystyle y_0=
\frac{J^2(1+ \sqrt{1+4\epsilon^2+2J^2})}{4\epsilon^2 + 2J^2},\\ \\
\displaystyle b_{\pm}=
\epsilon + {1\over 2} \pm \sqrt{(\epsilon + {1 \over 2})^2 - J^2},
\end{array}\end{equation}
\noindent
and

\begin{equation}\begin{array}{l}
x_1=\sqrt{b_-},\, x_2=\sqrt{b_+}, \\  \\
\sqrt{2\epsilon}< J < \epsilon + {1\over 2}, 
 \epsilon > {1\over 2}.
\end{array}\end{equation}
\noindent
As has been discussed in Ref.\cite{WKB}, the solutions
 (A11) and (A13) correspond to 
 the case when the classically allowed region extends into
the condensate and to the case when it 
is totally outside the condensate, respectively.

The $U,V$-amplitudes inside the classically allowed region are 
\cite{WKB} 

\begin{equation}\begin{array}{l}
\displaystyle 
U={C_0(\epsilon, J) \over 2r^{3/2}_c }
\left(S_++S_-\right){\sin\phi\over x\sqrt{v_x}} Y_{L,L_z},\\  \\
\displaystyle V= {C_0(\epsilon, J) \over 2r^{3/2}_c }
\left(S_+-S_-\right){\sin\phi\over x\sqrt{v_x}} Y_{L,L_z},
\end{array}\end{equation}
\noindent
Here $S_{\pm}=\sqrt{\sqrt{1+(k/\epsilon)^2} \pm (k/\epsilon)}$;
 $Y_{L,L_z}$ is the spherical harmonic;
the normalization constant is 

\begin{equation}
C_0^{-2}(\epsilon, J)={1\over 2}\int^{x_2}_{x_1}{dx\over v_x},
\end{equation}
\noindent
and the dimensionless radial velocity\\
 $v_x=\sqrt{mr_{ho}^2/\hbar
 \omega_{ho}
r_c^2}v_r$ is given by

\begin{equation}\begin{array}{r}
\displaystyle v_x= \sqrt{2{\epsilon^2+k^2 \over \epsilon^2}
(\sqrt{\epsilon^2+k^2} - k - {J^2 \over 2x^2})}, \\ \\
 x_1 <x \leq 1,
\end{array}\end{equation}
\noindent
inside the condensate and by 

\begin{equation}\begin{array}{r}
\displaystyle v_x= \sqrt{2\epsilon +1 - x^2 - J^2/x^2}, \\ \\
 1 < x < x_2
\end{array}\end{equation}
\noindent
outside the condensate.  In calculation of $C_0(\epsilon, J)$ 
and in what follows,
we replace $\sin^2\phi$ by $1/2$ because the WKB phase
 $\phi $ \cite{WKB} varies rapidly inside the classically
allowed region. The integrals outside this region are exponentially
small, and we neglect them.

Substituting (A14) into (A1) and employing the
 units (A9) in Eq.(34), we find
the expressions (35), (36) where the dimensionless function $D(\beta)$
is defined as
 
\begin{equation}
\displaystyle
D(\beta)=\left[\int_0^{\infty} d\epsilon 
{{\rm e}^{ \epsilon /\beta }\over \left({\rm e}^{\epsilon /\beta } -1\right)^2} 
\rho (\epsilon)\right]^{1/2},
\end{equation}
\noindent
with the notation  

\begin{equation}\begin{array}{l}
\displaystyle
\rho (\epsilon)= \int^{J_0(\epsilon)}_0 dJJ
{C^2_0(\epsilon, J)\over 2} \{
\int^{x_2}_{x_1}{dx\over v_x}
[\epsilon -\\  \\
\displaystyle
 ({1\over 2}x^2 - {1\over 2} + 2k)\sqrt{1+ {k^2\over \epsilon^2}}
+{ k^2 \over \epsilon}]\}^2
\end{array}\end{equation}
\noindent
introduced, and $k$ determined in Eq.(A10).
 The value of the limit $J_0 (\epsilon)$
can be found from Eqs.(A11), (A13). Specifically, for $\epsilon \leq 1/2$,
only the case (A11) can be realized. This implies that

\begin{equation}
J_0(\epsilon)= \sqrt {2\epsilon},
 \quad \epsilon \leq {1\over 2}.
\end{equation}
\noindent
For $\epsilon > 1/2$, Eq.(A13) yields 

\begin{equation}
J_0(\epsilon)= \epsilon + {1\over 2}, \quad \epsilon >{1\over 2}.
\end{equation}
\noindent
Consequently, the integral (A19) can be expressed as 
$\rho (\epsilon)= \rho_1 (\epsilon) + \rho_2 (\epsilon)$, 

\begin{equation}\begin{array}{l}
\displaystyle
\rho_1 (\epsilon)= \int^{\sqrt{2\epsilon}}_0 dJJ
{C^2_0(\epsilon, J)\over 2}[{\rm In}_1(\epsilon, J) + 
{\rm In}_2(1,\epsilon, J)]^2,\\ \\
\displaystyle
\rho_2 (\epsilon)= \\  \\ 
\displaystyle \Theta(\epsilon - {1\over 2})
\int^{\epsilon +1/2}_{\sqrt{2\epsilon}} dJJ
{C^2_0(\epsilon, J)\over 2} [{\rm In}_2(\sqrt{b_-},\epsilon, J)]^2, 
\end{array}\end{equation}
\noindent
where the notations 

\begin{equation}\begin{array}{l}
\displaystyle {\rm In}_1(\epsilon, J)=\int^{x_2}_{x_1}{dx\over v_x}{\epsilon 
\sqrt{\epsilon^2 + k^2} \over k + \sqrt{\epsilon^2 + k^2}}, \quad x_1 < 1;\\ \\
\displaystyle
{\rm In}_2(\alpha, 
\epsilon, J)=\int^{\sqrt{b_+}}_{\alpha}{dx\over v_x}(\epsilon
+{1\over 2} - {1\over 2}x^2), \quad \alpha \geq 1, 
\end{array}\end{equation}
\noindent
are introduced; $C^2_0(\epsilon, J)$ is given by Eq.(A15), and
$v_x$ is determined by Eqs.(A16), (A17). Note that here we have employed
the explicit expressions (A11)- (A13) for the turning points. 

The value of the normalization constant $C_0$ can be found explicitly
\cite{WKB}. The integrals (A23) can also be calculated explicitly.
We find

\begin{equation}
{C^2_0(\epsilon, J)\over 2}={2\over \pi}, \,\, 
{\rm In}_2(\sqrt{b_-}, \epsilon, J)= {\pi \over 4}(\epsilon + {1\over 2}),
\end{equation}
\noindent 
for $ \epsilon > 1/2,\,\,  \sqrt{2\epsilon} < J < \epsilon +1/2 $, and 

\begin{equation}\begin{array}{c}
\displaystyle {C^2_0(\epsilon, J)\over 2}=\left( {2\epsilon \arccos \alpha_1
\over \sqrt{2\epsilon^2 + J^2}} + \arccos \alpha_2 \right)^{-1}, \\ \\
\displaystyle
{\rm In}_1(\epsilon, J)=
 {\epsilon^2 \arccos \alpha_1\over \sqrt{2\epsilon^2 + J^2}}
+{1\over 4} \sqrt{2\epsilon - J^2} -\\ \\
\displaystyle {1\over 2\sqrt{2}} \ln \alpha_3,\\  \\
\displaystyle {\rm In}_2(1, \epsilon, J)= {1\over 2}(\epsilon + {1\over 2})
\arccos \alpha_2 -\\  \\
\displaystyle {1\over 4} \sqrt{2\epsilon - J^2},
\end{array}\end{equation}
\noindent
for $ J < \sqrt{2\epsilon}$, where we have introduced the notations

\begin{equation}\begin{array}{l}
\displaystyle \alpha_1=\sqrt{{2\epsilon^2 +J^2 - \epsilon
+\epsilon\sqrt{1+4\epsilon^2 + 2J^2} \over 2
 \epsilon\sqrt{1+4\epsilon^2 + 2J^2}}},\\ \\
 \displaystyle \alpha_2= \sqrt{{{1\over 2} - \epsilon +
\sqrt{(\epsilon +{1\over 2})^2 - J^2}\over 2
 \sqrt{(\epsilon +{1\over 2})^2 - J^2}}},\\ \\
\displaystyle \alpha_3= {1\over \sqrt{2}}[\sqrt{1+
2\epsilon - \sqrt{1+4\epsilon^2 + 2J^2}}
+\\  \\
\displaystyle \sqrt{ 1+ 2\epsilon +\sqrt{1+4\epsilon^2 + 2J^2}}]\\ \\
(1+4\epsilon^2 + 2J^2)^{-1/4}.
\end{array}\end{equation}
\noindent
We note that in the formal limit $\beta >> 1$, the function $D(\beta)$
given by (A18)
can be found explicitly. Indeed, in this case the main contribution to 
(A19) comes from $\epsilon >> 1$. This implies that only the term
 $\rho_2(\epsilon)$ 
in Eq.(A22) should be taken into account 
because it gives the highest power of
$\epsilon$ as $\rho_2(\epsilon)\sim \epsilon^4$. As simple
analysis of (A22) shows, the term $\rho_1(\epsilon)
\sim \epsilon^3$. Thus, taking 
$\rho (\epsilon) \approx \rho_2 (\epsilon)$ and 
combining Eqs. (A24), (A22), (A25) and (A12),      we find

\begin{equation}
\displaystyle \rho (\epsilon)= {\pi \over 16}(\epsilon^2 -
 {1\over 4})^2\Theta ( \epsilon
-{1\over 2})
\approx {\pi \over 16}\epsilon^4,
\end{equation}
\noindent
for $\epsilon >> 1$. Substituting this into (A18) and taking the limit
$\beta >>1$, we obtain 

\begin{equation}
D(\beta)\approx \sqrt{{3\pi \over 2}} \beta^{5/2},
\end{equation}
\noindent
which yields Eq.(38). This expression is shown in Fig.1 by the dashed line.
As one can see, in the range of $\beta $ of the order of 1 the
approximation (A28) underestimates the rate by approximately
20\%.

We note that actual values of $\beta =T/ 2\mu $
 are far from being $\beta >> 1$.
Indeed, employing Eqs. (A2) and (37), we find

\begin{equation}
\beta = {T\over 2\mu}= \left({r_{ho} \over 15a}\right)^{2/5}\zeta^{-1/3}(3)
N^{-1/15}{T\over T_c},
\end{equation}
\noindent
which yields values $\beta \approx  1$ for the experiment  \cite{DAMP1}
for $T \approx T_c$.
 Therefore, for these
values the function $D(\beta)$ should be found numerically
(see the solid line in Fig.1).

In the opposite limit $\beta \to 0$, which corresponds to
$T << 2\mu$ or large $N$, the contribution to $D(\beta)$ 
due to $\rho_2(\epsilon)$ becomes exponentially
small. Thus, the term $\sim \rho_1(\epsilon)$ (A22) dominates
in (A18). Taking
into account that the effective values of $\epsilon \sim \beta$,
one may perform an expansion in terms of the small
parameter $\epsilon$ in
Eqs.(A26), (A25) and obtain 

\begin{equation}
\rho(\epsilon) \approx \rho_1(\epsilon)= \rho_{01} \epsilon^{7/2},
\quad \epsilon \to 0,
\end{equation} 
\noindent
where the notation 

\begin{equation}\begin{array}{l}
 \rho_{01} =\\  \\
\displaystyle {1\over \sqrt{2}}\int^{\pi/2}_0
dx \sin x {[x + {1\over 12}\sin 2x (7+ 11\cos^2)]^2\over x + {1\over 2}\sin 2x}
\end{array}\end{equation}
\noindent
has been introduced.
 A numerical evaluation of this integral gives $\rho_{01} \approx 1.5$.
This yields for (A18)

\begin{equation}\begin{array}{c}
\displaystyle D(\beta)\approx D_0\beta^{9/ 4},\\ \\
D_0=\left[\rho_{01} \int^{\infty}_0dx{{\rm e}^x \over ( {\rm e}^x - 1)^2}
x^{7/2}\right]^{1/2} \approx 4.4
\end{array}\end{equation}
\noindent
in the limit $\beta << 1$. 

\section{Calculation of the amplitude dependence of the dephasing rate} 
Expanding Eq.(29) up to the third order with respect to $\eta$, one 
gets 

\begin{equation}
\ddot{\eta}+\omega^2_M\eta -\alpha_M\eta^2 + \beta_M\eta^3=0 ,
\end{equation}
\noindent
where 
the notations are

\begin{equation}\begin{array}{c}
\omega^2_M=(5-\xi)\omega^2_{ho},\quad
\alpha_M=10(1-{2\over 5}\xi)\omega^2_{ho},\\ \\
\beta_M=20(1-{\xi \over 2})\omega^2_{ho}.
\end{array}\end{equation}
\noindent
The solution of (B1) up to the second order with respect
to $\eta_0$ has
a form

\begin{equation}\begin{array}{c}
\displaystyle
\eta = {2\alpha_M \over \omega^2_M}|\eta_0|^2+
(\eta_0{\rm e}^{i\omega t} + c.c.) -\\  \\
\displaystyle
{\alpha_M \over 3 \omega^2_M}(\eta_0^2 {\rm e}^{i2\omega t} + c.c.), 
\end{array}\end{equation}
\noindent
where the effective frequency $\omega$ in the same order is

\begin{equation}
\omega = \omega_M + \omega'|\eta_0|^2, \quad
\omega'=-{5\alpha_M^2\over 3\omega^3_M} +{3\beta_M\over 2\omega_M}.
\end{equation}
\noindent
Now employing (B2) in (B4) and performing the thermal averaging 
of (B1) over $\xi$ in the limit $\xi << 1$, we obtain 

\begin{equation}
\displaystyle
\langle \eta\rangle_T = 
\eta_0 {\rm e}^{i\langle \omega \rangle_Tt
 - t^2/\tau_{dM}^2(A_1)} +c.c.,
\end{equation}
\noindent
where the constant as well as the second harmonic have been 
omitted; the dephasing rate $1/\tau_{dM}(A_1)$ as a function
of the amplitude  $A_1=2\eta_0$ of the first harmonic is given by Eq.(39).




\end{document}